\documentclass[onecolumn,preprintnumbers,11pt,amsmath,amssymb,nofootinbib]{revtex4}

\def	\be	{\begin{equation}}
\def	\ee	{\end{equation}}
\def	\bqt	{\begin{quote}}
\def	\eqt	{\end{quote}}

\begin{document}

\title{Implications of nonsymmetric metric theories for particle physics. New interpretation of the Pauli coupling}

\author{Gin\'{e}s R.P\'{e}rez Teruel$^1$}

\affiliation{$^{1}$ $^1$Departamento de F\'{i}sica Te\'{o}rica, Universidad de Valencia, Burjassot-46100, Valencia, Spain} 

\begin{abstract}
\begin{center}
{\bf Abstract}
\end{center}
\noindent
In this work we provide a possible geometrical interpretation of the spin of elementary particles. In particular, it is investigated how the wave equations of matter are altered by the addition of an antisymmetric contribution to the metric tensor. In this scenario the explicit form of the matter wave equations is investigated in a general curved space-time, and then the equations are particularized to the flat case. Unlike traditional approaches of NGT, in which the gravitational field is responsible for breaking the symmetry of the flat Minkowski metric, we find more natural to consider that, in general, the metric of the space-time could be nonsymmetric even in the flat case. The physical consequences of this assumption are explored in detail. Interestingly enough, it is found that the metric tensor splits into a bosonic and a fermionic; the antisymmetric part of the metric is very sensitive to the spin and turns out to be undetectable for spinless scalar particles. However, fermions couple to it in a non-trivial way (only when there are interactions). In addition, the Pauli coupling is derived automatically as a consequence of the nonsymmetric nature of the metric
\end{abstract}

\maketitle

\section{Introduction}
\label{introduction}

\thispagestyle{empty}

\noindent
One important motivation for a geometrical formulation of some properties of the elementary particles
originates from the following observation. The unification of fundamental interactions requires that we unify, preliminarly, two distinct theories: classical General Relativity and Quantum Mechanics. It is reasonable to expect that what will eventually emerge will entail a substantial revision of both theories, which we may expect, will emerge merely as approximations to some underlying new theory which encompasses them.\cite{Cle}
Other motivation for our geometrical formulation is the following. There is no physical principle that ensures that the metric tensor has to be symmetric, $g_{\mu\nu}=g_{\nu\mu}$; this condition is assumed a priori without a physical justification.
On the other hand, the extensions of General Relativity (GR) represent an active field of research with very different theoretical approximations since the formulation of the original theory in 1915. In the first years, several authors\cite{Ein1,En2,Edd}\cite{Sch,Sch2,Wey}, tried to unify electromagnetism and gravity by means of a formalism in which the metric tensor and the affine connection were not symmetric. These theories, despite their high degree of mathematical elegance did not work, and failed in the attempt to recover some classical results such us the Lorentz Force. 

In the last years, the discovery of the cosmic speedup and other cosmological issues such as the nature of dark matter have increased the theoretical interest in the possible generalizations of GR. In fact, among these attemps to go beyond Einstein's gravity, several authors\cite{Da,Moff,Je} have recovered the study of general theories based on nonsymmetric space-times, making some progress and achieving consistent classical field theories that avoid the problems that plagued the previous developments. We will accept the theoretical framework of NGT in Palatini formalism as our starting point, in order to investigate the implications for particle physics. In particular, we want to understand how the wave equations of the matter fields will be affected by the addition of a non symmetric contribution in the metric tensor.The mathematical discussion that follows provides some results that allow us to claim that the Pauli matrix, $\sigma_{\mu\nu}$, is proportional to the antisymmetric part of the metric, suggesting a possible deep and unexpected relation between spin and geometry. Indeed, we will see in detail how a nonsymmetric metric tensor can be decomposed in its bosonic and fermionic contributions.The standard equation that identifies the symmetric part of the metric with the anticommutator of the gamma matrices (Clifford algebra) is recovered in a natural way, while the commutator of the gamma matrices is found to be related with the antisymmetric (fermionic), part of the metric. In the light of this outcome, we will make the hypothesis that the metric is nonsymmetric not only in the presence of a gravitational field, but also in the flat case. This only assumption when applied to the electromagnetic interaction, provides a direct derivation of the Klein-Gordon-Pauli equation (KGP), wich is being recently considered as an alternative to the Dirac equation in QED \cite{La}\cite{La2}. This is motivated for the necessity of understanding experimental results such as the measurement of the muonic hydrogen Lamb shift. The discrepancy between Dirac-QED and experiment is reinterpreted as a revised size of the proton. In this work the KGP equation is directly deduced from the formalism, which allows us to give a theoretical justification of the Pauli coupling.
\section{General Overview of NGT in Palatini formalism}
\label{Palatini}

\noindent
The simplest NGT generalization of GR comes from the following Lagrangian \cite{Da}

\be\label{Lagrangian}
\displaystyle\mathcal{L}_{NGT}(g,\Gamma)=\sqrt{-g}g^{\mu\nu}R_{\mu\nu}(\Gamma)
\,
\ee

where $\Gamma^{\mu}_{\alpha\beta}$ is a priori independent from the metric (Palatini formalism), and should be obtained through the corresponding field equations. 

The nonsymetric space-time metric $g^{\mu\nu}$, can be decomposed in its symmetric and antisymmetric parts
\be\label{metric}
\displaystyle g_{\mu\nu}=G_{\mu\nu}+B_{\mu\nu}
\,
\ee
where
\be
G_{\mu\nu}\equiv g_{(\mu\nu)}=\frac{1}{2}\Big(g_{\mu\nu}+g_{\nu\mu}\Big)
\,
\ee
\be
B_{\mu\nu}\equiv g_{[\mu\nu]}=\frac{1}{2}\Big(g_{\mu\nu}-g_{\nu\mu}\Big)
\,
\ee

Indeed, there is no physical principle that guarantees that the metric should be symmetric, for this reason it is worth studying the consequences of this generalization. Regarding the affine connection, it is also nonsymmetric and admits a similar decomposition $\Gamma^{\sigma}_{\alpha\beta}\equiv  \Gamma^{\sigma}_{(\alpha\beta)}+\Gamma^{\sigma}_{[\alpha\beta]}$

 Varyng (\ref{Lagrangian}) with respect to the metric and the connection provides the field equations
\be
\partial_{\mu}g_{\alpha\beta}-\widetilde{\Gamma}^{\sigma}_{\alpha\mu}g_{\sigma\beta}-\widetilde{\Gamma}^{\sigma}_{\mu\beta}g_{\alpha\sigma}=0
\,
\ee
\be
R_{\alpha\beta}(\Gamma)-\frac{1}{2}g_{\alpha\beta}R=0
\,
\ee
Here, $R\equiv g^{\mu\nu}R_{\mu\nu}$ and the Ricci, $R_{\mu\nu}\equiv R^{\rho}_{\mu\rho\nu}=\partial_{\rho}\Gamma^{\rho}_{\mu\nu}-\partial_{\nu}\Gamma^{\rho}_{\mu\rho}+\Gamma^{\sigma}_{\mu\nu}\Gamma^{\rho}_{\sigma\rho}-\Gamma^{\sigma}_{\mu\rho}\Gamma^{\rho}_{\sigma\nu}$, depends on $\Gamma$ and has no a priori dependence on $g_{\mu\nu}$. We also have a recursive relation between $\widetilde{\Gamma}^{\sigma}_{\alpha\beta}$ and $\Gamma^{\sigma}_{\alpha\beta}$
\be
\widetilde{\Gamma}^{\sigma}_{\alpha\beta}=\Gamma^{\sigma}_{\alpha\beta}+\frac{2}{3}\Gamma_{\beta} \delta^{\sigma}_{\alpha} 
\,
\ee
With $\Gamma_{\mu}\equiv \frac{1}{2}\Gamma^{\lambda}_{[\mu\lambda]}$.These relations represent the pure gravitational sector of the theory. It is interesting to note that slight variations of the Lagrangian given in (\ref{Lagrangian}) yield to different theories\cite{Da,Moff,Je} that are able to explain interesting dark matter features such as the galactic rotation curves, assuming that the B-field is a massive field, satisfying a Proca type equation in the first order approximation \cite{Moff}. In spite of these differences, the general geometrical structure of all the NGT is basically the same.All these theories have in common a nonsymmetric metric tensor and the independent affine connection equipped with torsion. In the next subsection of this work, we will focus on the matter sector of these theories, rather than the purely gravitational,in order to investigate how the inclusion of the B-field into the metric affects the wave equations of matter

\section{Matter wave equations}
\label{Matter_Weq}

\noindent
In Minkowski space-time, given a symmetric metric tensor $g_{\mu\nu}=g_{\nu\mu}$ with signature (+ - - -) the first Casimir operator will acquire the following form for a test particle of four-momentum $P^{\mu}$:
\be\label{Casimir}
g_{\mu\nu}P^{\mu}P^{\nu}=P_{\nu}P^{\nu}=m^2
\ee
By application of the correspondence principle $P_{\mu}\rightarrow i\partial_{\mu}$ we obtain the free Klein-Gordon wave equation
\be\label{klein_curved}
\big(g_{\mu\nu}\partial^{\mu}\partial^{\nu}+m^2\big)\phi(x)=0
\ee
It is interesting to note that if we add an antisymmetric field $B_{\mu\nu}$ to the metric in (\ref{Casimir}) the Casimir and therefore the Klein-Gordon equation will not be affected by this addition.This is due to the fact that $P^{\mu}P^{\nu}$ does not couple to the B-field. Indeed, $P^{\nu}P^{\mu}$ is a purely symmetric object describing ordinary matter. However, when we have interactions, the correspondence principle is modified by inserting covariant derivatives $P_{\alpha}\rightarrow iD_{\alpha}$ instead of usual derivatives $P_{\alpha}\rightarrow i\partial_{\alpha}$. This substitution has the effect of breaking the symmetry of $P^{\alpha}P^{\beta}$, because $D^{\alpha}D^{\beta}$ no longer commutes, and this will generate an additional term involving the antisymmetric part of the metric.
To show this in detail, let us write the Klein-Gordon equation in curved nonsymmetric space-times
\be\label{curved_KG}
\big(g^{\alpha\beta}D_{\alpha}D_{\beta}+m^2\big)\phi(x)=\Big[g^{\alpha\beta}\Big(\frac{1}{2}\{D_\alpha,D_\beta\}+\frac{1}{2}[D_{\alpha},D_{\beta}]\Big)+m^2\Big]\phi(x)
\,
\ee
\\
Where, $D_{\alpha}D_{\beta}\equiv\frac{1}{2}\{D_{\alpha},D_{\beta}\}+\frac{1}{2}[D_{\alpha},D_{\beta}].$
Given the decomposition, $g^{\alpha\beta}\equiv G^{\alpha\beta}+B^{\alpha\beta}$, we obtain after a direct computation in the last equation the following result
\be\label{curved_KG2}
\Big(\frac{1}{2}G^{\mu\nu}\{D_\mu,D_\nu\}+\frac{1}{2}B^{\mu\nu}[D_{\mu},D_{\nu}]+m^2\Big)\phi(x)=0
\,
\ee
For scalar fields, $D_{\alpha}\phi(x)=\partial_{\alpha}\phi(x)$. Taking into account this identity, after some algebra we find the following expression for the commutator of the covariant derivatives
\be\label{Commutator}
[D_{\alpha},D_{\beta}]\phi(x)=-\Big(\Gamma^{\mu}_{\alpha\beta}-\Gamma^{\mu}_{\beta\alpha}\Big)\partial_{\mu}\phi(x)
\,
\ee
GR is torsion-free and this means that the connection (Levi-Civita) is symmetric. In these conditions the last commutator vanishes. Nevertheless, in our analysis from the point of view of NGT this term is no longer zero because the connection is not symmetric, and it will provide an additional contribution that we should take into account. 
In the same fashion, it can be found an expression for the anticommutator of the covariant derivatives, but involving the symmetric part of the affine connection
\be\label{anticommutator}
\{D_{\alpha},D_{\beta}\}\phi(x)=2\partial_{\alpha}\partial_{\beta}\phi(x)-\Big(\Gamma^{\mu}_{\alpha\beta}+\Gamma^{\mu}_{\beta\alpha}\Big)\partial_{\mu}\phi(x)
\,
\ee
Equations (\ref{Commutator},\ref{anticommutator}) can be used to write the compact form of the Klein-Gordon field in curved nonsymmetric space-times. Indeed, replacing the last results in (\ref{curved_KG2}) we obtain after straightforward manipulations
\be\label{curved_KG3}
\Big(G^{\alpha\beta}\partial_{\alpha}\partial_{\beta}+m^2\Big)\phi(x)=g^{\alpha\beta}\Gamma^{\mu}_{\alpha\beta}\partial_{\mu}\phi(x)
\ee
Where
\be
\Gamma^{\mu}_{\alpha\beta}=\frac{1}{2}\Big(\Gamma^{\mu}_{\alpha\beta}+\Gamma^{\mu}_{\beta\alpha}\Big)+\frac{1}{2}\Big(\Gamma^{\mu}_{\alpha\beta}-\Gamma^{\mu}_{\beta\alpha}\Big)
\,
\ee
The left side of Eq. (\ref{curved_KG3}) is identical to the corresponding Klein-Gordon equation in GR. The difference lies in the right side:now the metric and the connection are not symmetric, but it is worth noting that the form of the equation (in the minimal coupling, without additional terms such as curvature scalars) remains the same.

What about Dirac fields? The explicit and detailed treatment of Dirac fields in a general curved space-time is a much more complicated task (see for instance \cite{Wel}) but we only want to take a general picture in order to inquire some aspects of the field $B_{\mu\nu}$. For this reason, we will not solve the covariant derivative over Dirac fields. We will limit to the task of requiring that the Klein-Gordon equation could be expressed as the square of the Dirac equation. 

As is well known, the Dirac equation in curved space-time can be written as
\be\label{Dirac}
\Big(i\gamma^{\alpha}D_{\alpha}-m\Big)\Psi(x)=0
\,
\ee
Where, $D_{\alpha}\Psi(x)=\partial_{\alpha}\Psi(x)+\Gamma_{\alpha}\Psi(x)$. $\Gamma_{\alpha}$ includes the spin-connection but it is not necessary for our purpuses to write the explicit expression here. Now, we multiply the Dirac equation by the following operator
\be
\Big(-i\gamma^{\alpha}D_{\alpha}-m\Big)\Big(i\gamma^{\beta}D_{\beta}-m\Big)\Psi(x)=0
\,
\ee

If we assume $D_{\alpha}\gamma^{\beta}=0$ which seems a plausible generalization of the condition $\partial_{\alpha}\gamma^{\beta}=0$ that is verified by the Dirac matrices in a flat space-time, we find

\be\label{Dirac_1}
\Big(\gamma^{\alpha}\gamma^{\beta}D_{\alpha}D_{\beta}+m^2\Big)\Psi(x)=0
\,
\ee
This is nothing but the Klein-Gordon equation in general curved nonsymmetric space-times (\ref{curved_KG}).

Using the fact that $\gamma^{\alpha}\gamma^{\beta}=\frac{1}{2}\{\gamma^{\alpha},\gamma^{\beta}\}+\frac{1}{2}[\gamma^{\alpha},\gamma^{\beta}]$, we can write Eq. (\ref{Dirac_1}) as
\be
\Big(\frac{1}{4}\{\gamma^{\alpha},\gamma^{\beta}\}\{D_{\alpha},D_{\beta}\}+\frac{1}{4}[\gamma^{\alpha},\gamma^{\beta}][D_{\alpha},D_{\beta}]+m^2\Big)\Psi(x)=0
\,
\ee
\\
Finally, comparing this last expression with Eq. (\ref{curved_KG2}), we can make the identification
\be\label{Clifford1}
G^{\alpha\beta}=\frac{1}{2}\{\gamma^{\alpha},\gamma^{\beta}\}
\,
\ee
\be\label{Clifford2}
B^{\alpha\beta}=\frac{1}{2}[\gamma^{\alpha},\gamma^{\beta}]
\,
\ee
The identification of the symmetric part of the metric tensor with the anticommutator of the gamma matrices is a well known result of QFT (Clifford algebra). On the other hand, the commutator of the gamma matrices transforms as a tensor, and is a clue concept to understand the behaviour of the Dirac field under Local Lorentz Transformations. We suggest a new interpretation of this tensor in the framework of nonsymmetric space-times, where the metric tensor has an antisymmetric part.
We recall that in the previous derivation, it was assumed the condition, $D_{\alpha}\gamma^{\beta}=0$, which implies
\be
\partial_{\alpha}\gamma^{\beta}=-\Gamma^{\beta}_{\alpha\mu}\gamma^{\mu}
\,
\ee
This equation determines the \emph{parallel transport} of the vector field $\gamma^{\beta}$. A similar result can be found in the literature \cite{Hes1,Hes2,Pav}. It is possible to use this relation together with (\ref{Clifford1}) to build a theory of gravitation in Clifford Space, where the fundamental objects are the Dirac matrices $\gamma^{\beta}$ rather than the metric $g^{\alpha\beta}$\cite{Cas}.
It is important to note that in the context of Nonsymetric Gravitational Theories (NGT), the gravitational field itself is the agent who breaks down the symmetry of the original flat metric,which is symmetric. However, since the commutator of the gamma matrices is a well defined object that has a rich meaning in Minkowski space-time, it seems natural to assume that Eq. (\ref{Clifford2}) is completely general, even in the flat case, due to the fact that (\ref{Clifford1}) is also true in the flat case (Clifford Algebra). In other words, it seems more natural to think that the validity of (\ref{Clifford2}) implies that the metric tensor is in general nonsymmetric even in the flat case. In the next section we will exploit and study the consequences of this assumption.
\section{The Flat case. The Bosonic and Fermionic parts of the metric tensor}
\label{NEC-lower dim}

\noindent
Since the antisymmetric part of the metric has passed unnoticed until now, it must have very special properties. If we assume that Eqs. (\ref{Clifford1},\ref{Clifford2}) are also true for the Minkowski space-time, then a general nonsymmetric flat metric $g_{\mu\nu}\equiv\gamma_{\mu}\gamma_{\nu}$ can be decomposed as

\be
g_{\mu\nu}=G_{\mu\nu}+B_{\mu\nu}=\frac{1}{2}\{\gamma_{\mu},\gamma_{\nu}\}+\frac{1}{2}[\gamma_{\mu},\gamma_{\nu}]=\frac{1}{2}\{\gamma_{\mu},\gamma_{\nu}\}-i\sigma_{\mu\nu}
\ee
where
\be
\sigma_{\mu\nu}=\frac{i}{2}[\gamma_{\mu},\gamma_{\nu}]
\,
\ee
This is the Pauli matrix in 4 dimensions. We can therefore observe that the antisymmetric part of the metric is very sensitive to the spin. The last equation suggests that a nonsymmetric metric tensor may be decomposed in its bosonic and fermionic contributions. The first interesting consequence of such decomposition is that, for spinless escalar particles $\sigma_{\mu\nu}\phi=0$. This means that scalar particles, such as the Higgs, cannot couple to the antisymmetric part of the metric tensor. For this type of matter the antisymmetric part of the metric turns out to be undetectable, and the associated klein-Gordon equation in the presence of an interaction (such as the electromagnetic field), will be
\be
\Big(g^{\mu\nu}D_{\mu}D_{\nu}+m^2\Big)\phi=\Big(\frac{1}{2}G^{\mu\nu}\{D_{\mu},D_{\nu}\}+m^2\Big)\phi(x)=0
\,
\ee
\\
However, the situation for fermions is more subtle. As we discussed in Eqs. (\ref{Casimir},\ref{klein_curved}) a free particle will only couple to the symmetric part of the metric because $P_{\alpha}P_{\beta}$ is symmetric, due to the Poincare relations, $[P_{\mu},P_{\nu}]=0$ that obey the generators of space-time translations, a symmetry that fulfill both fermions and bosons. However, when we have interactions, $P_{\alpha}\rightarrow iD_{\alpha}$, which implies that $D_{\alpha}D_{\beta}$ no longer commutes and it will generate an additional contribution involving $\sigma_{\mu\nu}$. In this case, fermions will couple to the Pauli matrix $\sigma_ {\mu\nu}$, the antisymmetric part of the metric. Indeed, the associated Klein-Gordon equation for fermions will acquire the following form
\be\label{fermionic_0}
\Big(\widetilde{\Box}+m^2\Big)\Psi=\Big(\frac{1}{2}G^{\mu\nu}\{D_{\mu},D_{\nu}\}-\frac{i}{2}\sigma^{\mu\nu}[D_{\mu},D_{\nu}]+m^2\Big)\Psi=0
\,
\ee
With $\widetilde{\Box}\equiv g^{\mu\nu}D_{\mu}D_{\nu}$.
As we shall see, when this equation is particularized to the electromagnetic field, it reduces to the Klein-Gordon-Pauli equation (KGP), which is currently being utilized to solve precision-QED problems.\cite{La}.

On the other hand, if $\sigma_{\mu\nu}$ is the fermionic part of metric, then it should retain metric properties. The metric nature of $\sigma_{\mu\nu}$ is not evident when this tensor is coupled to Dirac spinors. However, over Rarita-Schwinger fields these metric properties are manifest as can be shown easily. Indeed, the Rarita-Schwinger field $\psi_{\rho}$ describes fermions of spin $3/2$ and satisfies the following equations
\begin{align}
(i\gamma^{\mu}\partial_{\mu}-m)\psi_{\rho}=0\nonumber\\
\gamma^{\nu}\psi_{\nu}=0
\end{align}
Where the second equation is a constraint to remove the undesired degrees of freedom of spin $1/2$. Multiplying this equation by $\gamma^{\mu}$ yields
\be\label{Rarita}
\gamma^{\mu}\gamma^{\nu}\psi_{\nu}=\Big(\frac{1}{2}\{\gamma^{\mu},\gamma^{\nu}\}+\frac{1}{2}[\gamma^{\mu},\gamma^{\nu}]\Big)\psi_{\nu}=G^{\mu\nu}\psi_{\nu}-i\sigma^{\mu\nu}\psi_{\nu}=0
\ee
Since $G^{\mu\nu}\psi_{\nu}=\psi^{\mu}$, Eq. (\ref{Rarita}) implies
\be\label{fermionic}
i\sigma^{\mu\nu}\psi_{\nu}=\psi^{\mu}
\ee
The tensor $i\sigma^{\mu\nu}$ has therefore the property of rising and lowering indexes when acting over Rarita-Schwinger fermions. Thus, if the metric tensor has an antisymmetric part, the identification of $\sigma^{\mu\nu}$ with the antisymmetric part of the metric seems natural and consistent following the results of Eqs. (\ref{Clifford1},\ref{Clifford2}) and (\ref{fermionic})
\subsection{The electromagnetic interaction. Derivation of the Pauli coupling}
\label{Pauli_coupling}

\noindent
We want to investigate the explicit form of Eq. (\ref{fermionic_0}) in the presence of the electromagnetic field. In this scenario the covariant derivative, $D_{\mu}\Psi(x)$, will be equal to
\be
D_{\mu}\Psi(x)=\Big(\partial_{\mu}+ieA_{\mu}\Big)\Psi(x)
\,
\ee
Where $e$ is the electric charge and $A_{\mu}$ the potential vector. Using this identity together with the definition of the electromagnetic field-strength $F_{\mu\nu}\equiv \partial_{\mu}A_{\nu}-\partial_{\nu}A_{\nu}$, it is possible to compute the commutator that appears in Eq. (\ref{fermionic_0})
\be
-\frac{i}{2}\sigma^{\mu\nu}[D_{\mu},D_{\nu}]\Psi(x)=\frac{e}{2}\sigma^{\mu\nu}F_{\mu\nu}\Psi(x)
\,
\ee
This is the Pauli coupling, a term introduced by hand in the Dirac equation in order to explain some experimental results\cite{Con}. It has been derived here as an inevitable consequence of the formalism. It is interesting to note that this term emerges automatically as long as one assumes the nonsymmetric nature of the Minkowski space-time metric. Regarding the contribution that contains the coupling of the symmetric tensors $G^{\mu\nu}\{D_{\mu},D_{\nu}\}$ we find
\newpage
\be
\frac{1}{2}G^{\mu\nu}\{D_{\mu},D_{\nu}\}\Psi=\Box\Psi-e^{2}A^{2}\Psi+\frac{ie}{2}G^{\mu\nu}F_{(\mu\nu)}\Psi+2ieA^{\mu}\partial_{\mu}\Psi
\,
\ee
We have defined, $\Box\Psi(x)\equiv G^{\mu\nu}\partial_{\mu}\partial_{\nu}\Psi(x)$. It should be understood that we are using the symmetric part of the metric to rise and lower indexes. In the last equation we also have defined the symmetric tensor, $F_{(\mu\nu)}\equiv \partial_{\mu}A_{\nu}+\partial_{\nu}A_{\mu}$. Notice that, $G^{\mu\nu}F_{(\mu\nu)}\Psi(x)=2\partial_{\mu}A^{\mu}\Psi(x)$, and this term gives no contribution if we work in the Lorentz gauge ($\partial_{\mu}A^{\mu}=0$).
Combining the last results, we can write the associated Klein-Gordon equation for fermions in the Lorentz gauge as
\be
\Box\Psi-e^{2}A^{2}\Psi+2ieA^{\mu}\partial_{\mu}\Psi+\frac{e}{2}\sigma^{\mu\nu}F_{\mu\nu}\Psi+m^2\Psi=0
\,
\ee
This is the Klein-Gordon-Pauli equation, wich has received increasing attention in the last months due to the attempts to explain experimental results such as the ``Proton Size Puzzle" \cite{La}\cite{La2}.This equation can be obtained from the following Lagrangian density
\be
\mathcal{L}=\bar\Psi\Big(\Pi^{2}-m^2-\frac{ge}{4}\sigma^{\mu\nu}F_{\mu\nu}\Big)\Psi
\,
\ee
With $\Pi=i\partial^{\mu}-eA^{\mu}$, the Hermitian momentum operator. Note that the theory developed in this work predicts a value for the gyromagnetic ratio $g=2$. Due to quantum corrections, we know that this value is slightly different from $2$. However, it must be taken into account that the analysis introduced here is purely relativistic. The most important thing is that the form of the Pauli term has been directly deduced, not introduced ad hoc in the Lagrangian as it is found in other works. How to explain the small corrections to the relativistic result $g=2$ in the framework of Nonsymmetric Metric Theories, will be the subject of future research \cite{Te}
\subsection{Tachyons}
\label{Tachyons}

\noindent
As we have shown in the previous section, in a flat space-time spinless bosons do not couple to the antisymmetric part of the metric because, $\sigma_{\mu\nu}\phi(x)=0$. Meanwhile fermions couple to it only when there are interactions; interactions generate a coupling given by the term $\sigma^{\mu\nu}[D_{\mu},D_{\nu}]\Psi$. One naturally wonders whether it could exist some type of matter able to feel the antisymmetric part of the metric even in the free case. The answer to this question is positive (from a theoretical point of view), although this kind of matter would have some special properties incompatible with the Poincare symmetry. As is well known, all elementary particles fall in representations of this group, whose associated Lie algebra is given by the equations
\begin{align}\label{Poincare}
[P_{\mu},P_{\nu}]&=0\nonumber\\
\frac{ 1 }{ i }[M_{\mu\nu}, P_\rho]& = G_{\mu\rho} P_\nu - G_{\nu\rho} P_\mu\nonumber\\
\frac{ 1 }{ i }[M_{\mu\nu}, M_{\rho\sigma}] &= G_{\mu\rho} M_{\nu\sigma} - G_{\mu\sigma} M_{\nu\rho} - G_{\nu\rho} M_{\mu\sigma} + G_{\nu\sigma} M_{\mu\rho}
\,
\end{align}
Where $G_{\mu\nu}$ is the symmetric metric of Minkowski space-time, $P^{\mu}$ the generators of space-time translations and $M_{\mu\nu}$ the corresponding generators of Lorentz transformations.
We want to inquire the implications of relaxing any of the above commutative relations, in particular the first one.To this purpose, let us begin writing the left hand side of the Casimir (3.1) in a flat nonsymmetric space-time. Given a nonsymmetric metric tensor $g^{\mu\nu}\equiv\gamma^{\mu}\gamma^{\nu}=\frac{1}{2}\{\gamma^{\mu},\gamma^{\nu}\}+\frac{1}{2}[\gamma^{\mu},\gamma^{\nu}]$ in the free case (total absence of any interaction) we will have
\be\label{Tachyon}
\gamma^{\mu}\gamma^{\nu}P_{\mu}P_{\nu}=\frac{1}{4}\{\gamma^{\mu},\gamma^{\nu}\}\Big(P_{\mu}P_{\nu}+P_{\nu}P_{\mu}\Big)+\frac{1}{4}[\gamma^{\mu},\gamma^{\nu}]\Big(P_{\mu}P_{\nu}-P_{\nu}P_{\mu}\Big)
\,
\ee
It is clear from the above equation that a non zero coupling to the antisymmetric part of the metric in the case free would require a violation of the Poincare symmetry. Indeed, if matter has the ordinary properties given by the first commutator of (\ref{Poincare}), then the second term of the right hand side of (\ref{Tachyon}) will vanish,since $P_{\mu}P_{\nu}=P_{\nu}P_{\mu}$. However, let us investigate the theoretical consequences of such a violation of the Poincare symmetry. We are going to assume that certain exotic matter has an associated four-momentum $\hat{P_{\mu}}$ formed by a set of Grassmann numbers that verify
\be\label{Tachyon2}
\{\hat{P}_{\mu},\hat{P}_{\nu}\}=0
\,
\ee
Thus
\be\label{Tachyon3}
g^{\mu\nu}\hat{P}_{\mu}\hat{P}_{\nu}=\gamma^{\mu}\gamma^{\nu}\hat{P}_{\mu}\hat{P}_{\nu}=-\gamma^{\mu}\gamma^{\nu}\hat{P}_{\nu}\hat{P}_{\mu}=-m^2
\,
\ee
Where we have used the Dirac equation in momentum space, $\gamma^{\nu}\hat{P}_{\nu}\Psi=m\Psi$. The minus sign shows that we are dealing with tachyons. If we repeat the same reasoning with the usual four-momentum of the ordinary matter that fulfills the Poincare symmetry, a substitution in (\ref{Tachyon3}) will provide the correct sign for the square of the mass in the Casimir operator. 
This result arguably means that we are not allowed to treat fermions using Grassmann variables in the external space-time,but only in their internal, spinorial space. In any case we have demonstrated that free matter that couples to the antisymmetric part of the metric violates the Poincare symmetry and behaves like tachyons. The only acceptable coupling to the antisymmetric part of the metric takes place when there are interactions, as we have studied in detail in the previous sections.
\section{SUMMARY AND CONCLUSIONS}
\label{Tachyons}

\noindent
At the moment it is not known any physical principle that guarantees that the metric tensor of the space-time has to be symmetric. For this reason it is worth to investigate this class of theories from a theoretical point of view. The general idea was born long time ago in the context of GR and their extensions, such as classical unified field theories, and has been recovered successfully in the last years \cite{Da,Moff,Je}. We have reviewed the main aspects of these NGT, focusing on the matter sector instead of the pure gravitational sector. In contrast to the traditional approach, it was found that there is no reason to believe that the gravitational field is the responsible of breaking the symmetry of the original symmetric metric of Minkowski space-time. If the metric tensor is not symmetric, it seems more natural to consider that is not symmetric not only in the presence of a gravitational field, but also in the flat space-time. On the other hand, recent experiments in particle physics show that some important theoretical concept is missing. In this work we have shown that the inclusion of an antisymmetric part in the metric tensor provides a consistent framework in which the Klein-Gordon-Pauli equation is automatically derived, including the Pauli term as a direct consequence. It is interesting to note that the Pauli coupling has no theoretical justification in the usual approach of Electrodynamics where this term is always introduced ad hoc in the wave equations. In this sense, the decomposition of the metric into the symmetric (bosonic) and the antisymmetric (fermionic) parts, allow to understand the spin of the elementary particles in a geometrical manner, with the antisymmetric part of the metric undetectable for bosons in the flat case.

\end{document}